# VALIDITY OF KINETIC THEORY FOR RADIATIVE HEAT TRANSFER IN NANOPARTICLE CHAINS

Eric J. Tervo[1*], Baratunde A. Cola[1], Zhuomin M. Zhang[1]

[1]Georgia Institute of Technology, Atlanta, GA 30332, USA

ABSTRACT. In chains of closely-spaced nanoparticles supporting surface polaritons, near-field electromagnetic coupling leads to collective effects and super-Planckian thermal radiation exchange. Researchers have primarily used two analytical approaches to calculate radiative heat transfer in these systems: fluctuational electrodynamics, which directly incorporates fluctuating thermal currents into Maxwell's equations, and a kinetic approach where the dispersion relation provides modes and propagation lengths for the Boltzmann transport equation. Here, we compare results from the two approaches in order to identify regimes in which kinetic theory is valid and to explain differing results in the literature on its validity. Using both methods, we calculate the diffusive radiative thermal conductivity of nanoparticle chains. We show that kinetic theory is valid and matches predictions by fluctuational electrodynamics when the propagation lengths are greater than the particle spacing.

## 1. INTRODUCTION

When two objects are separated by distances smaller than the characteristic thermal wavelength, the net radiative heat flux may be significantly increased and exceed the far-field limit [1, 2]. This near-field enhancement is caused by the coupling of evanescent waves between objects, which exist due to total internal reflection of photons at the interfaces (frustrated modes) or due to surface modes including surface plasmon polaritons (SPPs) and surface phonon polaritons (SPhPs) [3, 4]. When surface modes are spatially confined to nanostructures such as nanoparticles they are considered "localized" SPPs or SPhPs. However, when many such nanostructures are placed close together, coupling of evanescent waves leads to delocalization and energy transport through the array of nanostructures [5-9], which has applications in sub-diffractional waveguiding [10-12] and thermal energy transport [13, 14]. It is therefore important to be able to accurately calculate thermal radiation exchange in these systems.

To predict the radiative heat transfer in linear chains of nanoparticles, two analytical methods have emerged. The first is an exact formalism in which fluctuating thermal currents are incorporated into Maxwell's equations as the source of thermal radiation, referred to as fluctuational electrodynamics (FED) [1]. FED has been extended to treat radiation exchange in systems of multiple small bodies, called many-body thermal radiation or the coupled dipole method [15-24]. The second method is based on a kinetic theory (KT) approach; the resonant, propagating modes are treated as energy carriers with a group velocity and propagation length determined from the complex dispersion [7, 8, 13, 25, 26].

In previous studies, we compared the spectral thermal emission from particle chains via FED to the density of states via the dispersion relation (corresponding to KT). Very good agreement was found between the two methods in predicting the peak shapes and locations [27]. We also calculated the radiative thermal conductivity and found that KT underpredicted the total conductivity but accounted for the dominant contributions [28]. On the other hand, a recent study by Kathmann *et al.* found that KT significantly overpredicted the radiative heat transfer near the resonance frequency and had additional limitations for metal nanoparticles supporting SPPs [29]. Here, we reconcile these differences by performing calculations with both methods for different materials and particle spacings.

---
[*] Corresponding Author: eric.tervo@gatech.edu.

## 2. THEORY AND METHODS

Let us consider a chain of isotropic, nonmagnetic nanoparticles of relative permittivity $\varepsilon(\omega)$ immersed in a non-absorbing medium of relative permittivity $\varepsilon_m$, as shown in Figure 1. The chain length is much greater than the propagation length of the surface polaritons, such that the heat transfer is diffusive. All particles are at temperature $T$ or $T + \Delta T$ with $\Delta T$ approaching zero as required to evaluate thermal conductance in the FED formalism.

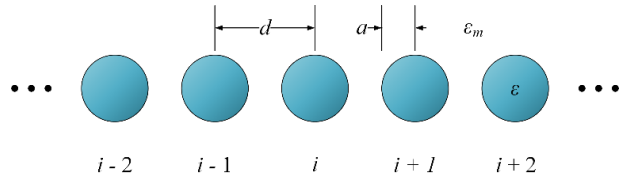

Figure 1. Schematic of nanoparticle chain under consideration.

The particles are spherical with radius $a = 25$ nm and are modeled as point dipoles, which is valid when the center-to-center spacing $d \gtrsim 3a$ [30-32]. The temperature is $T = 500$ K. The material is either hBN in $\varepsilon_m = 1$ as examined by Kathmann *et al.* [29] or SiC in $\varepsilon_m = 4$, because this permittivity maximizes interparticle coupling and corresponds with our previous work [27, 28]. The optical properties of both materials are described by a Lorentz model

$$\varepsilon = \varepsilon_\infty \left( 1 + \frac{\omega_{LO}^2 - \omega_{TO}^2}{\omega_{TO}^2 - \omega^2 - i\omega\Gamma} \right) \quad (1)$$

where $\varepsilon_\infty = 4.88$ (6.7), $\omega_{LO} = 3.032 \times 10^{14}$ ($1.82 \times 10^{14}$) rad s$^{-1}$, $\omega_{TO} = 2.575 \times 10^{14}$ ($1.49 \times 10^{14}$) rad s$^{-1}$, and $\Gamma = 1.001 \times 10^{12}$ ($8.92 \times 10^{11}$) rad s$^{-1}$ for hBN [29, 33] (SiC [34]). The dielectric functions are assumed to be independent of temperature. We note that although hBN is optically anisotropic, isotropic optical properties based on the ordinary component of the dielectric function tensor are used to match the analysis by Kathmann *et al.*

### 2.1 Fluctuational Electrodynamics Approach

The application of FED to $N$ bodies was pioneered by Ben-Abdallah and co-workers during the past decade [15, 17]. When each particle is modeled as a point dipole, the fields and polarizations at each particle can be expressed in a self-consistent set of equations using dyadic Green's functions and particle polarizabilities. If the particles cannot be modeled as dipoles, they may be discretized using the thermal discrete dipole approximation [20, 21] or they may be modeled with a multipolar framework [35]. The sources of the fields are fluctuating thermal currents whose correlation is given by the fluctuation dissipation theorem [36, 37]. In this framework, the power absorbed by particle $i$ is [17]

$$P_i = \int_0^\infty \frac{\hbar\omega}{2\pi} \sum_{j \neq i} \frac{4\chi_i\chi_j}{|\alpha_i|^2} n_{ji} \text{Tr}\left(\mathbb{T}_{ij}^{-1}\mathbb{T}_{ij}^{-1\dagger}\right) d\omega \quad (2)$$

where $\chi_i = \text{Im}(\alpha_i) - |\alpha_i|^2 k^3 (6\pi\varepsilon_m)^{-1}$, $k$ is the wavevector given by $\omega\sqrt{\mu_0\varepsilon_0\varepsilon_m}$, $\alpha_i$ is the dressed polarizability of particle $i$, $n_{ji} = f_{BE}(\omega, T_j) - f_{BE}(\omega, T_i)$ with $f_{BE}(\omega, T_i)$ the Bose-Einstein distribution at temperature $T_i$, and $\mathbb{T}_{ij}^{-1}$ is a subset of the inverted interaction matrix accounting for influences between all particles [17]. The dressed and Clausius-Mossotti polarizabilities are

$$\alpha_i = \left[\frac{1}{\alpha_i^{CM}} - \frac{ik^3}{6\pi\varepsilon_m}\right]^{-1} \quad (3)$$

and

$$\alpha_i^{CM} = 3\varepsilon_m V_i \frac{(\varepsilon_i - \varepsilon_m)}{(\varepsilon_i + 2\varepsilon_m)} \quad (4)$$

By noting that each term in the summation accounts for the heat transfer from particle $j$ to $i$, we can write the thermal conductance between two particles as

$$G_{ij} = \frac{\partial P_{ij}(T)}{\partial T} = \int_0^\infty \frac{\hbar\omega}{2\pi} \frac{4\chi_i\chi_j}{|\alpha_i|^2} \frac{\partial f_{BE}}{\partial T_i} \text{Tr}\left(\mathbb{T}_{ij}^{-1}\mathbb{T}_{ij}^{-1\dagger}\right) d\omega \quad (5)$$

To evaluate the diffusive thermal conductivity $\kappa$, we assume a very small, linear temperature gradient along the chain, such that the definition for conductance in Equation (5) may be applied between any two particles. The total heat flux crossing an imaginary plane between particles $i$ and $(i + 1)$ includes contributions from $i$ to $(i + 1)$, $i$ to $(i + 2)$, $(i - 1)$ to $(i + 1)$, $(i - 1)$ to $(i + 2)$, etc., and the contribution to the heat flux from each pair decreases as the distance between them increases. From this pattern, we can devise the equation [28]

$$\kappa = \frac{1}{S}\sum_{j=1}^{N} j G_{ij} L_{ij} \quad (6)$$

where $S = \pi a^2$ is the cross-sectional area of the chain, there are $N$ particles on each side of the center particle $i$, $j$ runs from the nearest to the farthest particle on one side of $i$, and $L_{ij}$ is the center-to-center distance between particles $i$ and $j$. $N$ is then increased until $\kappa$ converges. We use $N = 70$ for our calculations and have checked that this is sufficient in all cases via a maximum convergence criterion of 1% for consecutive iterations.

## 2.2 Kinetic Approach

In the second method, the energy carried by delocalized surface polaritons propagating along the chain is quantified and used to calculate the thermal conductivity. This approach was also first explored by Ben-Abdallah and co-authors [7, 8] in the 2000s. First, the complex dispersion for the propagating modes must be obtained [11]. There are two degenerate transverse ($\perp$) modes and one longitudinal ($\parallel$) mode, and their dispersion relations are, respectively [38]

$$0 = 1 + \frac{\alpha}{4\pi\varepsilon_m d^3}\left\{\left[\text{Li}_3 e^{i\left(\frac{\omega}{v}+k\right)d} + \text{Li}_3 e^{i\left(\frac{\omega}{v}-k\right)d}\right]\right.$$
$$\left. - \frac{i\omega d}{v}\left[\text{Li}_2 e^{i\left(\frac{\omega}{v}+k\right)d} + \text{Li}_2 e^{i\left(\frac{\omega}{v}-k\right)d}\right] - \left(\frac{\omega d}{v}\right)^2\left[\text{Li}_1 e^{i\left(\frac{\omega}{v}+k\right)d} + \text{Li}_1 e^{i\left(\frac{\omega}{v}-k\right)d}\right]\right\} \quad (7)$$

and

$$0 = 1 - \frac{\alpha}{2\pi\varepsilon_m d^3}\left\{\left[\text{Li}_3 e^{i\left(\frac{\omega}{v}+k\right)d} + \text{Li}_3 e^{i\left(\frac{\omega}{v}-k\right)d}\right] - \frac{i\omega d}{v}\left[\text{Li}_2 e^{i\left(\frac{\omega}{v}+k\right)d} + \text{Li}_2 e^{i\left(\frac{\omega}{v}-k\right)d}\right]\right\} \quad (8)$$

where $k$ is the wavevector, $v = c/\sqrt{\varepsilon_m}$ with $c$ the speed of light in vacuum, and $\text{Li}_n(x)$ is the polylogarithm function of order $n$. Either $\omega$ or $k$ must be taken as complex, and the dispersion may be solved by varying real $k$ over the first Brillouin zone $\pi/d$ and solving for complex frequency $\widetilde{\omega}$ or by varying real $\omega$ and solving for complex wavevector $\tilde{k}$. Once the dispersion is determined, the group velocity $v_g = \partial\omega/\partial Re(\tilde{k})$ ($v_g = \partial Re(\widetilde{\omega})/\partial k$) and propagation length $\Lambda = [2\text{Im}(\tilde{k})]^{-1}$ ($\Lambda = |v_g|[-2\text{Im}(\widetilde{\omega})]^{-1}$) may be found for the choice of $\tilde{k}$ ($\widetilde{\omega}$). For $\widetilde{\omega}$, the imaginary part will have negative solutions [11, 38]. This is because the dispersion relations assume a frequency dependence of $e^{-i\omega t}$, requiring $\text{Im}(\widetilde{\omega}) < 0$ for oscillations to decay as $t \to \infty$. The typical dispersion relations consequently contain diverging sums [11], which motivates the use of polylogarithms [39] as shown here.

Once the dispersions are solved, the diffusive thermal conductivity may be derived from the Boltzmann transport equation and reads [13, 29]

$$\kappa = \frac{1}{\pi S} \int_{\omega_{min}}^{\omega_{max}} \hbar\omega(2\Lambda_\perp + \Lambda_\parallel) \frac{\partial f_{BE}}{\partial T} d\omega \qquad (9)$$

where $\omega_{min}$ and $\omega_{max}$ represent the extent of the dispersion relation and $\Lambda_\perp$ and $\Lambda_\parallel$ represent the transverse and longitudinal propagation lengths, respectively.

## 3. RESULTS

With these equations in hand, we have the tools needed to calculate thermal conductivity with both the FED and KT approaches. First, we present representative dispersion relations and propagation lengths for both SiC and hBN nanoparticle chains, and then we present results for spectral thermal conductivity with different approaches for various nanoparticle spacings.

**3.1 Dispersions and Propagation Lengths**

Dispersion relations for SiC at $d = 3a$ and hBN at $d = 4a$ are shown in Figure 1(a) and (b), and the corresponding propagation lengths are shown in Figure 1(c) and (d). The calculations are performed for both $\widetilde{\omega}$ (solid lines) and $\tilde{k}$ (dashed lines). We also show the particle spacing in Figure 1(c) and (d), which will become important when considering the validity of KT.

Both dispersions exhibit the expected behavior and agree with previous calculations for SiC [27] and hBN [29]. For the choice of $\widetilde{\omega}$, there is strong coupling with radiative modes at the light line, especially for the transverse modes [11, 38] as illustrated by the insets. This leads to long propagation lengths near that frequency, which will have negligible impact on the radiative thermal conductivity due to the very small bandwidth. The dispersion and propagation lengths for $\tilde{k}$ agree fairly well with those for $\widetilde{\omega}$ in the regions of high group velocity, but at the edges of the dispersion there is substantial disagreement. In particular, the dispersion for $\tilde{k}$ diverges at large wavevectors and predicts low, consistent propagation lengths away from the resonance. The dispersion for $\widetilde{\omega}$ does not predict any modes in these regions. We also note that for the hBN chain, none of the propagation lengths reach the interparticle spacing, while for the SiC chain most of the longitudinal modes and the center of the transverse modes do. For hBN, this discrepancy calls into question the use of KT, because propagating modes cannot exist if they do not travel between adjacent particles [27, 28].

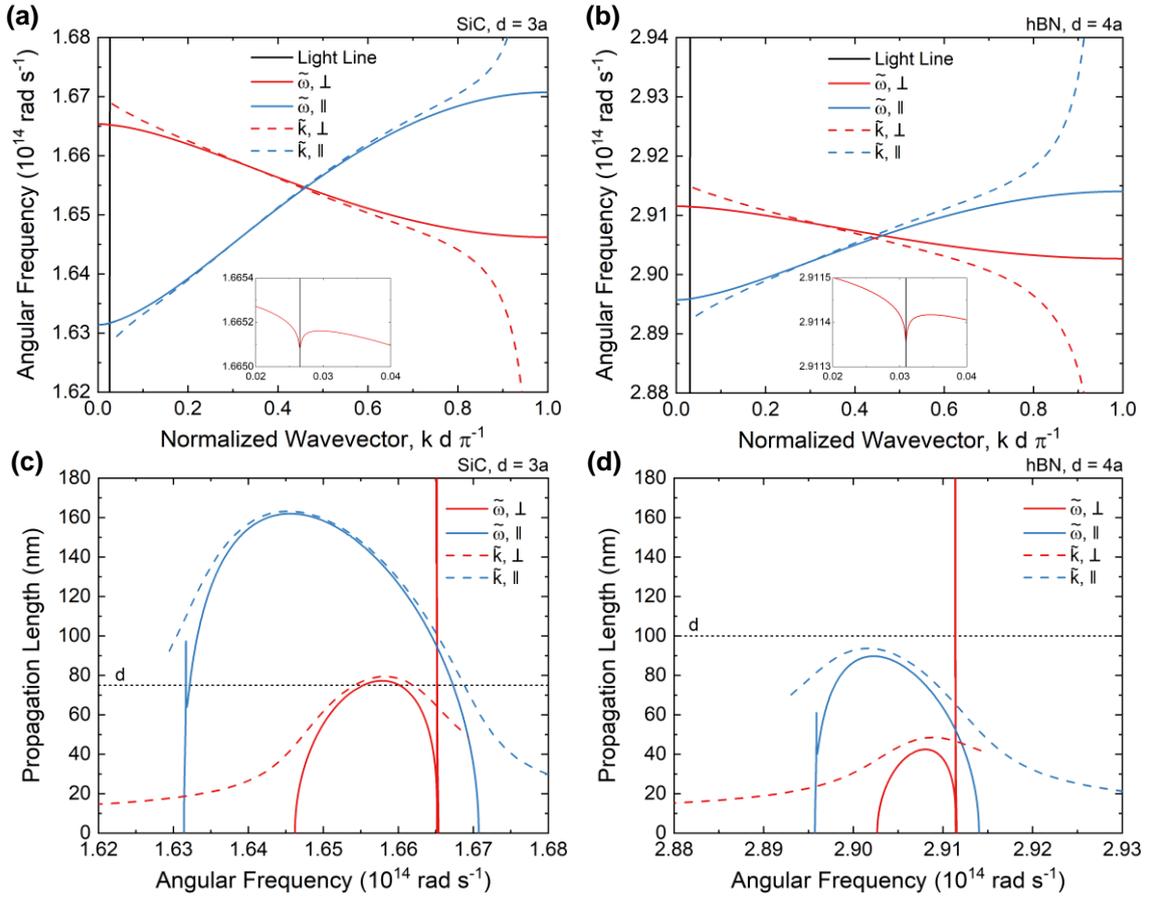

Figure 2. (a,b) Dispersion relations and (c,d) propagation lengths for (a,c) SiC and (b,d) hBN nanoparticle chains calculated with both complex $\omega$ and complex $k$.

### 3.2 Radiative Thermal Conductivity

We calculate the spectral radiative thermal conductivity via FED and KT with Equations (5), (6), and (9) before integrating, and we show the results for SiC at $d = 3a$ and hBN at $d = 4a$ in Figure 3(a) and (b).

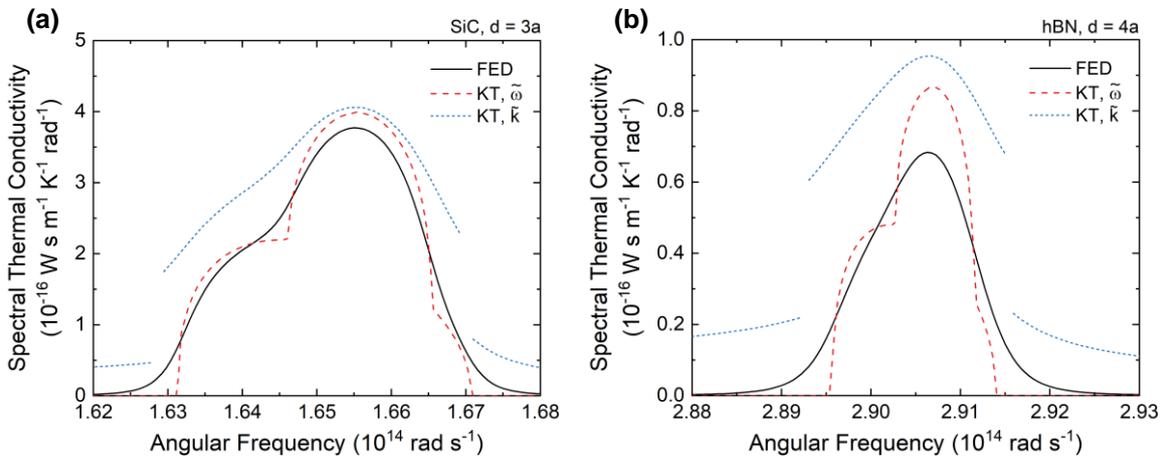

Figure 3. Spectral radiative thermal conductivity calculated with FED and KT approaches for (a) SiC and (b) hBN nanoparticle chains.

Comparing Figure 3(a) to our prior study [28], here we obtained the same results for KT with $\tilde{k}$. We obtained FED results which differ from our previous work by a factor of $2/\pi$, which was due to the use of an inappropriate form of the fluctuation dissipation theorem and has been corrected here. Our results in Figure 3(b) match those of Kathmann *et al.* [29] for FED and for KT with $\tilde{\omega}$ when their results are appropriately scaled by $\partial f_{BE}/\partial T$ for the difference in temperature and Fourier's law is used to calculate thermal conductivity. For the SiC case, there is generally good agreement between KT with $\tilde{\omega}$ and FED, with an average absolute error of 10.5% for KT where it is nonzero. The error for total thermal conductivity using KT with $\tilde{\omega}$ after integration is only $-3.4\%$ in comparison to the result from FED ($\kappa = 9.99 \times 10^{-4}$ W m$^{-1}$ K$^{-1}$). For hBN, on the other hand, KT with $\tilde{\omega}$ substantially overpredicts FED for the transverse (right peak) modes and has translational error for both the transverse and longitudinal modes. This agrees with the findings of Kathmann *et al.* When KT is used with $\tilde{k}$, the calculations overpredict FED for both materials over the whole spectrum, although the error is more pronounced for hBN.

To understand why KT performs better for SiC at $d = 3a$ than for hBN at $d = 4a$, we plot in Figure 4 the spectral thermal conductivity calculated with all methods for both materials at spacings of $d = 2a$ and $d = 5a$. The spacing of $2a$ violates the assumptions for the dipole approximation, so additional multipolar effects will cause the actual radiative thermal conductivity to differ from that shown here. However, the dipolar contributions to the radiative thermal conductivity predicted by the FED and KT approaches may be safely compared to each other, as they use the same assumptions.

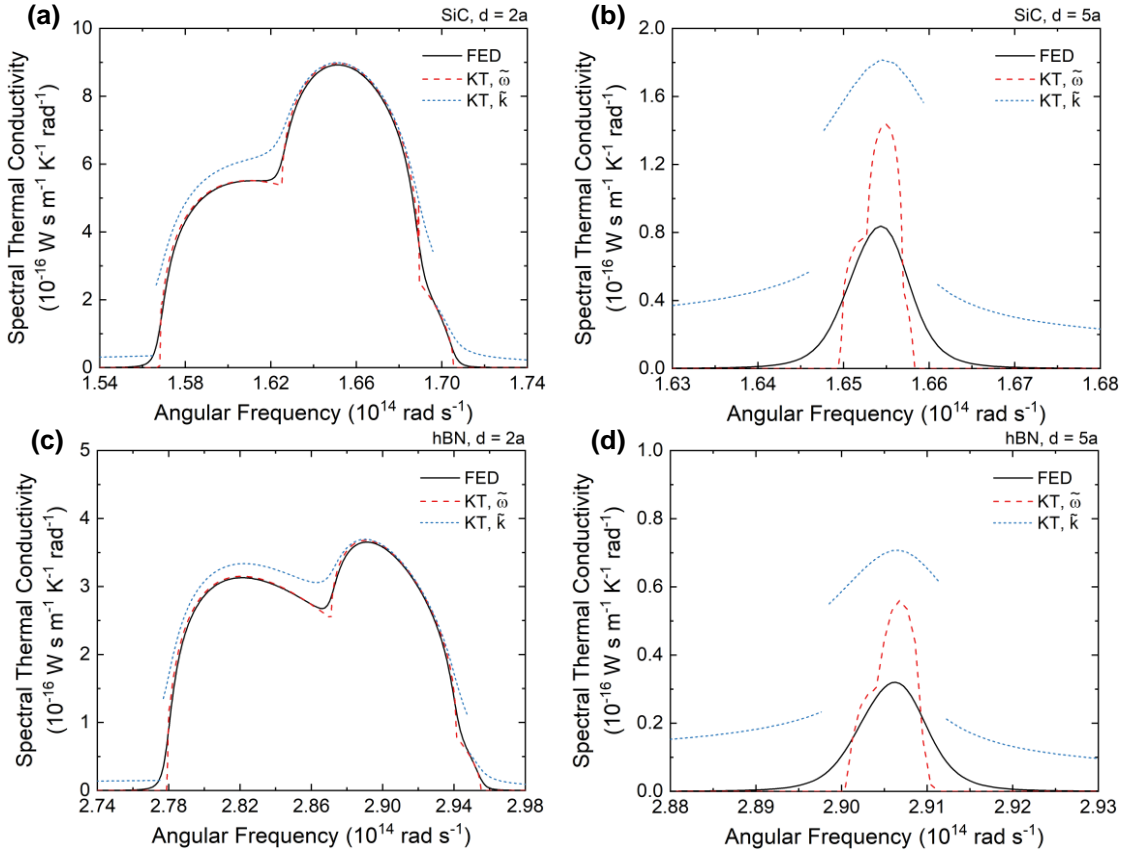

Figure 4. Spectral radiative thermal conductivity for (a,b) SiC and (c,d) hBN nanoparticle chains with separation distances of (a,c) $d = 2a$ and (b,d) $d = 5a$.

For the spacing of $d = 2a$ in Figure 4(a) and (c), we observe remarkable agreement between KT with $\tilde{\omega}$ and FED for both SiC and hBN. The average absolute error between the two curves is 3.3% (2.7%) for SiC (hBN) where the KT curve is nonzero. The error in total thermal conductivity is $-2.5\%$ ($0\%$) for SiC (hBN) compared to $\kappa = 0.0081$ W m$^{-1}$ K$^{-1}$ ($0.0048$ W m$^{-1}$ K$^{-1}$) from FED. KT with $\tilde{k}$ still overpredicts the spectral thermal conductivity for both materials at $d = 2a$, but the error is much less than the previous cases. On the other hand, at $d = 5a$ in Figure 4(b) and (d) KT clearly fails to predict both the shape and magnitude of the spectral thermal conductivity for both $\tilde{\omega}$ and $\tilde{k}$. The reason for this difference, as mentioned earlier, is how the propagation lengths compare to $d$. We plot in Figure 5 the propagation lengths for SiC at both spacings to illustrate this effect.

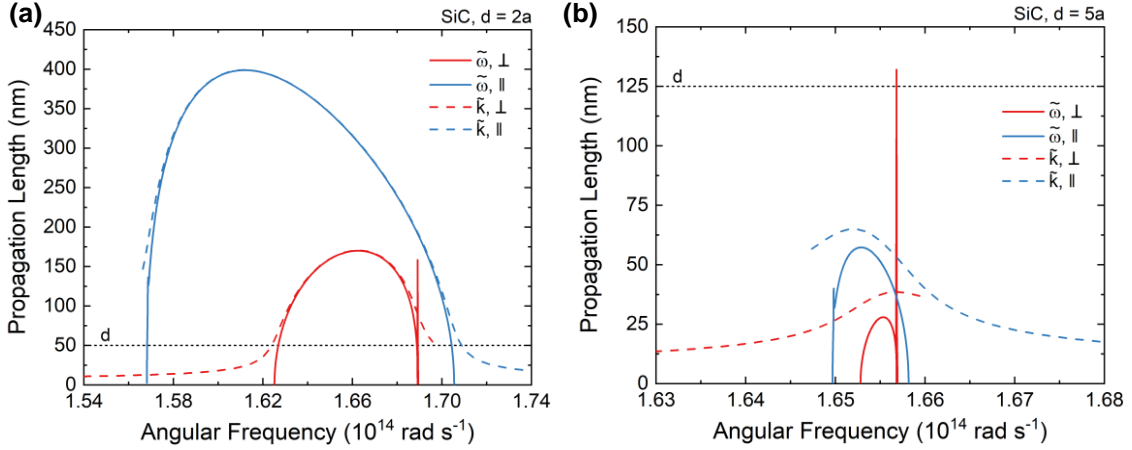

Figure 5. Propagation lengths for SiC chains at (a) $d = 2a$ and (b) $d = 5a$.

As expected, at closer spacing the stronger coupling leads to much longer propagation lengths. In the case shown in Figure 5(a), nearly all the modes predicted by KT with $\tilde{\omega}$ exceed $d$, and the maximum propagation lengths are about $8d$ for the longitudinal and about $3.4d$ for the transverse modes. It is also interesting to note that $\tilde{k}$ has much better agreement with $\tilde{\omega}$ at small spacing, which suggests it may be appropriate to use $\tilde{k}$ in these cases as long as the modes with $\Lambda < d$ are discounted. At farther spacing as shown in Figure 5(b), neither the transverse nor the longitudinal modes come close to $d$. Since the modes cannot span adjacent particles, it is not surprising that KT is an inappropriate formalism for these cases and gives inaccurate results as shown in Figure 4(b) and (d).

To further illustrate that KT corresponds with FED when the propagation lengths span adjacent particles, we perform calculations for the same system of hBN particles at $d = 4a$ as shown in Figure 3(b), but we reduce the damping coefficient from $\Gamma = 1.001 \times 10^{12}$ rad s$^{-1}$ to $\Gamma = 1 \times 10^{11}$ rad s$^{-1}$. This fictional material has much lower loss, so the transverse modes now propagate up to about $4d$ and the longitudinal modes up to about $9d$ as shown in Figure 6(b). Just as with the closely-spaced SiC and hBN particle chains, the FED and KT approaches agree very well as shown in Figure 6(a), especially when $\tilde{\omega}$ is used with KT.

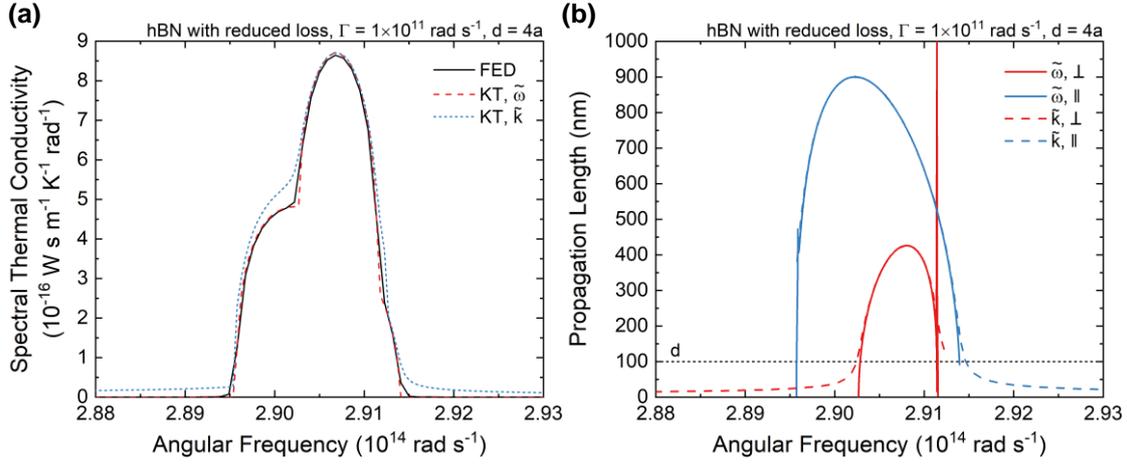

Figure 6. (a) Spectral radiative thermal conductivity and (b) propagation lengths for hBN at $d = 4a$ with a reduced damping coefficient $\Gamma = 1\times10^{11}$ rad s$^{-1}$.

## 6. CONCLUSION

The success of KT with $\widetilde{\omega}$ in regimes of strong coupling where $\Lambda > d$ demonstrates conclusively that KT can be a valid method to predict heat transfer by propagating surface polaritons in linear chains of nanostructures. This contradicts the conclusions of Kathmann *et al.* [29] because they only examined chains where the majority of the modes had $\Lambda < d$, and this illustrates that care must be taken to use KT only in its regime of validity. We also showed that KT with $\widetilde{k}$ tends to overpredict KT with $\widetilde{\omega}$, especially at high and low frequencies where $\Lambda < d$. The overprediction decreased when the coupling strength and $\Lambda$ increased, however, showing that KT with $\widetilde{k}$ could still lead to good estimates in these cases if used carefully.

Despite these results, KT still has several limitations when compared to FED as pointed out by Kathmann *et al*. When the polaritonic resonance lies in the frequency range that is thermally excited, such as with SiC and hBN, these contributions will dominate the radiative heat transfer. When the resonance is at much higher frequency, such as with noble metals, these modes are relatively unpopulated and a method such as FED is required to account for radiative transfer at lower frequencies. Another drawback of KT is that it cannot be readily applied to magnetic contributions to heat transfer, which may be important contributors in certain systems.

KT can, however, be a useful tool in some scenarios where it is difficult to apply FED. One of the reasons for the success of KT in studying phonon transport, for example, is that phonon dispersions may be calculated or experimentally determined and used as an input to KT [40]. Polaritonic dispersion relations for nanostructure arrays can also be simulated [5, 41] or accessed experimentally [42], permitting KT to make heat transfer predictions from these data. Additionally, material structures in anisotropic or nonhomogeneous environments [12] would be difficult to analyze with FED but are much more accessible with KT. We therefore expect KT to continue to be a useful tool in the analysis of heat transfer by propagating surface modes.

## ACKNOWLEDGEMENT

E.J.T. acknowledges support by the National Science Foundation Graduate Research Fellowship Program under Grant No. DGE-1650044. Z.M.Z. would like to thank the support from the U.S. Department of Energy, Office of Science, Basic Energy Sciences (DE-SC0018369).